\documentstyle[12pt]{article}
\begin{document}
\title{Bohmian trajectories and quantum phase space distributions}
\author{Nuno Costa Dias\footnote{{\it nuno.dias@ulusofona.pt}} \\ Jo\~{a}o Nuno Prata\footnote{{\it joao.prata@ulusofona.pt}} \\ {\it Departamento de Matem\'atica} \\
{\it Universidade Lus\'ofona de Humanidades e Tecnologias} \\ {\it Av. Campo Grande, 376, 1749-024 Lisboa, Portugal}}

\maketitle
\begin{abstract}
We prove that most quasi-distributions can be written in a form similar to that of the de Broglie-Bohm distribution, except that ordinary products are replaced by some suitable non-commutative star product. In doing so, we show that the Hamilton-Jacobi trajectories and the concept of "classical pure state" are common features to all phase space formulations of quantum mechanics. Furthermore, these results provide an explicit quantization prescription for classical distributions.
\end{abstract}
{\it PACS:} 03.65.Ca; 03.65.Db; 03.65.Ge \\
{\it Keywords:} De Broglie-Bohm interpretation, quasi-distributions, Cohen's classification

\section{Introduction}

In addition to the standard canonical and path integral quantizations there is an alternative method for quantizing classical theories, namely {\it deformation quantization} \cite{Lee}-\cite{Flato}. The virtue of this method resides in its formal representation which, in many respects, mimics that of classical statistical mechanics. A phase-space {\it quasi-distribution} \cite{Lee}, which describes the state of the system, plays the role of the wave function. Instead of operators, observables are represented by ordinary functions in phase space. With hindsight, the results hitherto obtained in this context can be summarized in a nutshell: quantum mechanics is obtained from classical statistical mechanics by replacing the ordinary abelian product of functions by some appropriate non-commutative associative {\it star product}. To be more precise, the previous quantization prescription permits a complete specification of all the relevant objects of phase space quantum mechanics with (thus far) the notorious exception of the quasi-distribution.

The arbitrariness in the choice of star product is related to ordering ambiguities of operators in quantum mechanics. We may choose e.g. to symmetrize operator products completely (Weyl order \cite{Weyl}), or else to normal order them (creation operators precede annihilation operators). To this myriad of ordering prescriptions there is a corresponding number of star products and quasi-distributions. In his work, Cohen \cite{Cohen} has sought to construct a general formalism where all possible quasi-distributions (and, consequently, all star products) are related by a sort of gauge transformation which leaves physically relevant quantities (mean values, probabilities) unscathed.

Another alternative formulation of quantum mechanics is the de Broglie-Bohm theory which displays a set of remarkable properties that are not shared by the quasi-distribution formulation \cite{Holland}-\cite{Bohm} namely that it is based on a positive defined distribution and that it allows for a causal interpretation in terms of the so-called "Bohmian trajectories". Both the de Broglie-Bohm and the quasi-distribution formulation attempt to provide a "classical-like" description of quantum mechanics. The two formulations are obviously equivalent as their predictions are identical to those of standard operator quantum mechanics. This notwithstanding, it does not diminish the conceptual and practical interest of each individual formulation. Firstly, because in some cases it may be advantageous to perform a specific calculation in one formulation rather than another, and secondly, because they provide conceptual insights in the interpretation of quantum phenomena.

In this letter we show that some key features of the de
Broglie-Bohm interpretation are shared by the quasi-distribution
formulation, enforcing the idea that these features play a key
role in the foundations of the quantum theory (previous work
studying the relation between the two formulations, but exploring
different points of view can be found in
\cite{Holland2,Takabayasi,Leavens,Polavieja,Dias3}). We will prove
that: 1) the dynamics of all quasi-distributions (and in
particular of the Wigner function) can be cast in terms of the de
Broglie-Bohm causal trajectories, 2) the classical pure state is
the semiclassical limit of all pure state quasi-distributions, 3)
all quasi-distributions are $*$-deformations of a classical pure
state.

From a more technical point of view, these results yield new calculation tools that may find interesting applications in a wide range of specific problems:
1) they constitute a more intuitive and explicit realization of Cohen's classification, providing a straightforward relation between a particular star-product and the corresponding quasi-distribution, 2) they formulate the dynamics of all quasi-distributions in terms of a system of two second order partial differential equations, 3) they provide an alternative implementation of the Wigner-Weyl map for density matrices, 4) they yield a (previously missing) quantization prescription for classical distributions.

\section{Quasi-distributions}

We will consider for simplicity one-dimensional systems. Let $\psi (x,t)$ be the wave function representing the state of the system at any given time $t$. The wave function is a solution of the Schr\"odinger equation, with Hamiltonian $\hat H = \frac{\hat p^2}{2m} + V(\hat x)$. Let $\hat A (\hat x, \hat p )$ be any observable, with a complete orthonomal set of eigenvectors $\left\{ |a> \right\}$ associated with the eigenvalues $\left\{a \right\}$\footnote{We will consider for simplicity nondegenerate spectra, although this is not crucial. The spectra can be continuous or discrete.}. The average value of $\hat A$ is evaluated according to: $< \hat A ( \hat x, \hat p)> = \int dx \hspace{0.2 cm} \psi^* (x) \hat A \left( x, - i \hbar \frac{d}{d x} \right) \psi (x)$, where we assume that $| \psi (x)|^2 $ is properly normalized. Moreover the probability for the measurement of the observable $\hat A$ yielding the eigenvalue $a$ is given by:
\begin{equation}
|< \psi | a>|^2 = \left| \int dx \hspace{0.2 cm} \psi^* (x) \psi_a (x)  \right|^2,
\end{equation}
where $\psi_a (x) = <x|a>$ is the eigenvector $|a>$ in the position representation.

Alternatively, the previous postulates of quantum mechanics can be reformulated in terms of phase-space quasi-distributions. According to Cohen's classification \cite{Cohen}, the $f$-quasi-distribution $F^f (x,p,t)$ for a pure state is given by:
\begin{equation}
F^f (x,p,t) = \frac{1}{4 \pi^2} \int d \xi \int d \eta \int d x' \hspace{0.2 cm} \psi^* \left( x' - \frac{\hbar}{2} \eta ,t \right) \psi \left( x' + \frac{\hbar}{2} \eta ,t \right) f (\xi , \eta , t) e^{i \xi (x' - x) - i \eta p},
\end{equation}
where $f (\xi, \eta ,t)$ is an arbitrary analytic function of $\xi$ and $\eta$. For notational convenience, we will henceforth omit the time dependence. In terms of $F^f$ the average value of $\hat A$ is given by the very appealing formula:
\begin{equation}
< \hat A ( \hat x, \hat p)> = \int dx \int dp \hspace{0.2 cm} F^f (x,p) A_f (x,p),
\end{equation}
where $A_f (x,p)$ is a c-number known as the "$f$-symbol" associated with the operator $\hat A$:
\begin{equation}
A_f (x,p) = \frac{\hbar}{2 \pi} \int d \xi \int d \eta \hspace{0.2 cm} Tr \left\{\hat A (\hat x, \hat p) e^{i \xi \hat x + i \eta \hat p} f^{-1} (\xi, \eta) \right\} e^{- i \xi x - i \eta p}.
\end{equation}
The "$f$-map" $V_f : \hat{\cal A} ({\cal H}) \to {\cal A} (T^*M)$ attributes to any operator $\hat A$ in the quantum algebra $\hat{\cal A}$, acting on the Hilbert space ${\cal H}$, the corresponding $f$-symbol in the "classical" algebra ${\cal A}$ over the phase space $T^*M$. It can be viewed as a Lie algebra isomorphism, provided ${\cal A} (T^*M)$ is endowed with a suitable Lie algebraic structure. Indeed, the $f$-map renders $T^* M$ into a non-commutative Lie algebra with an associative $*$-product and a $f$-bracket according to:
\begin{equation}
\left\{
\begin{array}{l l}
V_f \left( \hat A \cdot \hat B \right) = & A_f (x,p) *_f B_f (x,p)\\
& \\
V_f \left( \left[\hat A , \hat B \right] \right) = & \left[A_f (x,p), B_f (x,p) \right]_f = A_f  *_f B_f - B_f *_f A_f
\end{array}
\right.
\end{equation}
We could equally define a "dual $f$-symbol" for the operator $\hat A$:
\begin{equation}
\tilde V_f \left( \hat A (\hat x , \hat p) \right) = \tilde A_f (x,p) = \frac{\hbar}{2 \pi} \int d \xi \int d \eta \hspace{0.2 cm} Tr \left\{\hat A (\hat x, \hat p) e^{i \xi \hat x + i \eta \hat p} f (\xi, \eta) \right\} e^{- i \xi x - i \eta p}.
\end{equation}
In particular, the quasi-distribution $F^f$ is proportional to the dual $f$-symbol of the density matrix: $F^f = \frac{1}{2 \pi \hbar} \tilde V_f (\hat{\rho}) = \frac{1}{2 \pi \hbar} \tilde V_f (| \psi>< \psi|)$ \cite{Lee}. We can obviously define a dual $*$-product and a dual $f$-bracket, according to:
\begin{equation}
\left\{
\begin{array}{l l}
\tilde V_f \left( \hat A \cdot \hat B \right) = & \tilde A_f (x,p) *_f' \tilde B_f (x,p)\\
& \\
\tilde V_f \left( \left[\hat A , \hat B \right] \right) = & \left[\tilde A_f (x,p), \tilde B_f (x,p) \right]_f' = \tilde A_f  *_f' \tilde B_f - \tilde B_f *_f' \tilde A_f
\end{array}
\right.
\end{equation}
We shall call the sector of observables associated with the star product $*_f$ and the $f$-map $V_f$, the {\it observable's sector}. Likewise, the sector of the quasi-distribution, associated with the dual star product $*_f'$ and the dual $f$-map $\tilde V_f$ will be designated by {\it dual sector}.

There is an arbitrariness in the choice of function $f (\xi , \eta)$, or otherwise stated, in the choice of quasi-distribution $F^f$ and the corresponding $f$-symbols \cite{Lee,Cohen}. The physically relevant results such as eq.(1) remain unaltered. This constitutes the "gauge" invariance mentioned in the introduction.

It is worth assembling all the formulae concerning a change from a function $f_1 (\xi, \eta)$ to another function $f_2 (\xi, \eta)$ (or, from a quasi-distribution $F^1 (x,p)$ to another quasi-distribution $F^2 (x,p)$) \cite{Lee}:
\begin{equation}
\left\{
\begin{array}{l}
F^1 (x,p) = f_1 \left( i \frac{\partial}{\partial x} , i \frac{\partial}{\partial p} \right) f^{-1}_2 \left( i \frac{\partial}{\partial x} , i \frac{\partial}{\partial p} \right) F^2 (x,p)\\
\\
A_1 (x,p) = f^{-1}_1 \left( - i \frac{\partial}{\partial x} , - i \frac{\partial}{\partial p} \right) f_2 \left( - i \frac{\partial}{\partial x} , - i \frac{\partial}{\partial p} \right) A_2 (x,p)\\
\\
\tilde A_1 (x,p) = f_1 \left( i \frac{\partial}{\partial x} , i \frac{\partial}{\partial p} \right) f^{-1}_2 \left( i \frac{\partial}{\partial x} , i \frac{\partial}{\partial p} \right) \tilde A_2 (x,p)\\
\\
A_1 (x,p) *_1 B_1 (x,p)=  f^{-1}_1 \left( - i \frac{\partial}{\partial x} , - i \frac{\partial}{\partial p} \right) f_2 \left( - i \frac{\partial}{\partial x} , - i \frac{\partial}{\partial p} \right) A_2 (x,p) *_2 B_2 (x,p)\\
\\
\tilde A_1 (x,p) *'_1 \tilde B_1 (x,p)=  f_1 \left( i \frac{\partial}{\partial x} , i \frac{\partial}{\partial p} \right) f^{-1}_2 \left(  i \frac{\partial}{\partial x} ,  i \frac{\partial}{\partial p} \right) \tilde A_2 (x,p) *'_2 \tilde B_2 (x,p)
\end{array}
\right.
\end{equation}
The most celebrated example corresponds to the self-dual case $f_W (\xi , \eta )=1$, also known as the Wigner-Weyl formulation of quantum mechanics. The $*$-product (the Groenewold $*$-product \cite{Groenewold}) and the corresponding bracket (the Moyal bracket \cite{Moyal}) are given by:
\begin{equation}
\begin{array}{l l l}
V_W \left( \hat A \cdot \hat B \right)  & = A_W *_W B_W & = A_W (x, p) e^{\frac{i \hbar}{2} {\hat{\cal J}}}  B_W (x, p) ,\\
& & \\
V_W \left( \left[ \hat A , \hat B \right] \right)  & = \left[ A_W , B_W \right]_M & = 2 i  A_W (x, p) \sin \left(\frac{\hbar}{2} {\hat{\cal J}} \right) B_W ( x, p),
\end{array}
\end{equation}
where ${\hat{\cal J}}$ is the "{\it Poisson}" operator: $
{\hat{\cal J}} \equiv   \frac{ {\buildrel { \leftarrow}\over\partial}}{\partial q} \frac{ {\buildrel { \rightarrow}\over\partial}}{\partial p} -  \frac{{\buildrel {\leftarrow}\over\partial}}{\partial p}  \frac{{\buildrel { \rightarrow}\over\partial}}{\partial q}  $, the derivatives ${\buildrel { \leftarrow}\over\partial}$ and ${\buildrel { \rightarrow}\over\partial}$ acting on $A_W$ and $B_W$, respectively. The corresponding quasi-distribution is the celebrated Wigner function \cite{Wigner}. We shall discuss some of its properties in section 4.

Finally, to obtain the counterpart of eq.(1), we first need to introduce the $*$-delta function, which is a non-commutative generalization of the ordinary delta function \cite{Dias2,Flato}:
\begin{equation}
\delta^f_* \left( A(x,p) \right) = \frac{1}{2 \pi} \int d k \hspace{0.2 cm} e_{*_f}^{ik A (x,p)}.
\end{equation}
The $*$-exponential is defined by: $e_{*_f}^{B(x,p)} = \sum_{n=0}^{\infty} \frac{1}{n!} \left[B(x,p)\right]^{n*_f}$, where $\left[B(x,p)\right]^{n*_f}$ is the $n$-fold $f$-star product of $B(x,p)$. The $*$-delta function is a solution of the $*$-genvalue equation \cite{Fairlie1, Fairlie2}, i.e: $A_f (x,p) *_f \delta^f_* \left( A_f (x,p) - a \right) = \delta^f_* \left( A_f (x,p) - a \right) *_f A_f (x,p) = a \delta^f_* \left( A_f (x,p) - a \right)$, where $A_f = V_f \left(\hat A \right)$. The stargenvalue equation can be obtained upon application of the $f$-map to the eigenvalue equation: $\hat A|a><a| = a |a><a|$. In the Wigner-Weyl representation the $*$-delta function is a $\hbar$-deformation of the ordinary delta function \cite{Dias2}:
\begin{equation}
\delta^W_* \left( A (x,p) \right) = \delta \left( A (x,p) \right) -\frac{\hbar^2}{8} \Theta_1 (x,p) \delta''\left( A (x,p) \right) - \frac{\hbar^2}{24} \Theta_2 (x,p) \delta''' \left( A (x,p) \right)+ {\cal O} (\hbar^4),
\end{equation}
where $\Theta_1 (x,p) = \frac{\partial^2 A}{\partial x^2} \frac{\partial^2 A}{\partial p^2} - \left(\frac{\partial^2 A}{\partial x \partial p} \right)^2$ and $\Theta_2 (x,p) = \frac{\partial^2 A}{\partial x^2} \left( \frac{\partial A}{\partial p} \right)^2 - 2 \frac{\partial^2 A}{\partial x \partial p} \frac{\partial A}{\partial x} \frac{\partial A}{\partial p} + \frac{\partial^2 A}{\partial p^2} \left( \frac{\partial A}{\partial x} \right)^2$. In particular, we get for the position and momentum variables: $\delta^W_* \left(x - x_0 \right) = \delta \left(x-x_0 \right)$ and $\delta^W_* \left(p - p_0 \right) = \delta \left(p-p_0 \right)$.

\vspace{0.3 cm}
\noindent
We are now in a position to write a formula for $|< \psi|a>|^2$ in the phase space formulation\footnote{Here $A$ denotes generically the observable $\hat A (\hat x, \hat p)$ irrespective of the particular representation, i.e. irrespective of the choice of function $f(\xi, \eta)$.}:
\begin{equation}
{\cal P} \left( A=a \right) = \int dx \int dp \hspace{0.2 cm} F^f (x,p) \delta^f_* \left( A_f (x,p) -a \right).
\end{equation}

\section{De Broglie-Bohm interpretation}

In the De Broglie-Bohm interpretation of quantum mechanics \cite{Holland}-\cite{Bohm}, the wave function is written in the form:
\begin{equation}
\psi (x,t) = R(x,t) \exp \left( \frac{i}{ \hbar} S(x,t) \right),
\end{equation}
where $R(x,t)$ and $S(x,t)$ are some real functions. Substituting this expression in the Schr\"odinger equation we obtain the dynamics of $R$ and $S$:
\begin{equation}
\left\{
\begin{array}{l}
\frac{\partial {\cal P}}{\partial t} +  \frac{\partial}{\partial x} \left(\frac{{\cal P}}{m} \frac{\partial S}{\partial x}\right) =0 ,\\
\\
\frac{\partial S}{\partial t} + \frac{1}{2m} \left(\frac{\partial S}{\partial x} \right)^2 + V(x) - \frac{\hbar^2}{4m} \left[ \frac{1}{{\cal P}}
\frac{\partial^2 {\cal P}}{\partial x^2} - \frac{1}{2} \frac{1}{{\cal P}^2} \left( \frac{\partial {\cal P}}{\partial x} \right)^2 \right] =0,
\end{array}
\right.
\end{equation}
where ${\cal P} (x) \equiv |\psi (x)|^2 = R^2 (x)$, is the probability distribution. The first equation is a statement of probability conservation, with associated flux $\frac{{\cal P}}{m} \frac{\partial S}{\partial x}$. The second equation is interpreted as a Hamilton-Jacobi equation. The solution $S(x,t)$ represents an ensemble of trajectories, known as Bohmian trajectories, for particles under the influence of a classical potential $V(x)$ and a quantum potential:
\begin{equation}
Q(x,t) \equiv - \frac{\hbar^2}{2m} \frac{1}{R} \frac{\partial^2 R}{ \partial x^2} = - \frac{\hbar^2}{4m} \left[ \frac{1}{{\cal P}}
\frac{\partial^2 {\cal P}}{\partial x^2} - \frac{1}{2} \frac{1}{{\cal P}^2} \left( \frac{\partial {\cal P}}{\partial x} \right)^2 \right].
\end{equation}
Moreover, the momenta $p$ of the particles are subject to the constraint: $p = \frac{\partial S}{\partial x}$. The ensemble of Bohmian trajectories stem from the position $x$ at time $t=0$ with momentum $p= \frac{\partial S}{\partial x} (x,0)$ and probability ${\cal P} (x,0) = R^2 (x,0)$. Consequently, an ensemble of Bohmian trajectories can be described statistically by the phase-space distribution function \cite{Holland}:
\begin{equation}
F^B (x,p,t) \equiv R^2(x,t) \delta \left( p - \frac{\partial S}{\partial x} (x,t) \right),
\end{equation}
its dynamics being governed by the equation:
$$
\frac{\partial F^B}{\partial t} (x,p,t) = \left\{ \frac{p^2}{2m} + V(x) + Q(x,t) , F^B (x,p,t) \right\}_P,
$$
where $\left\{ , \right\}_P$ is the Poisson bracket.

\section{$*$-deformed causal form of quasi-distributions}

In this section we study some common features of the de Broglie-Bohm and the quasi-distribution formulations as well as their relation with classical statistical mechanics. For simplicity we shall provisionally focus on the self-dual Wigner-Weyl case ($f_W(\xi, \eta , t) = 1$).

The Wigner function \cite{Wigner} associated with $\psi (x,t)$ is given by:
\begin{equation}
F^W (x,p,t) \equiv \frac{1}{2 \pi} \int_{- \infty}^{+ \infty} dy e^{-i y p} \psi^* \left( x- \frac{1}{2} \hbar y, t \right) \psi \left( x+ \frac{1}{2} \hbar y, t \right).
\end{equation}
From the Schr\"odinger equation and eq.(17), it is possible to obtain the differential equation that dictates the dynamics of $F^W$:
\begin{equation}
\frac{\partial F^W}{\partial t} = \frac{1}{i \hbar} \left[H^W, F^W \right]_M ,
\end{equation}
where $H^W(x,p) = \frac{p^2}{2m} + V(x)$ is the Weyl symbol of the quantum Hamiltonian. The Wigner function is a real function and admits the marginal distributions:
\begin{equation}
{\cal P}(x) \equiv \int_{-\infty}^{+ \infty} dp F^W(x,p) = |\psi (x)|^2, \qquad
{\cal P}(p) \equiv \int_{-\infty}^{+ \infty} dx F^W(x,p) = |\phi (p)|^2,
\end{equation}
where $\phi (p)$ is the Fourier transform of $\psi(x)$\footnote{This result is a consequence of the fact that $\delta_*^W(x-x_0) = \delta (x-x_0)$ and $\delta_*^W(p-p_0) = \delta (p-p_0)$.}. From eq.(19) one could be tempted to interpret the Wigner function as a true probability distribution in phase space. However, this interpretation is immediately spoiled, if one realizes that it can take on negative values. Furthermore, and contrary to what happens in the de Broglie-Bohm formulation, the dynamics of the Wigner function does not, in general, allow for an interpretation in terms of causal trajectories (except for quadratic potentials).

We will now show that the Wigner distribution (17) and the de Broglie-Bohm distribution (16) are in fact an $\hbar$-deformation of each other. To do this it will prove useful to expand the Wigner function in powers of $\hbar$:
\begin{equation}
\begin{array}{c}
F^W (x,p,t) = \frac{1}{2 \pi} \sum_{n=0}^{\infty} \sum_{m=0}^{\infty} \frac{(-1)^n}{n! m!} \left( \frac{\hbar}{2} \right)^{n+m}
\frac{\partial^n \psi^*}{\partial x^n} \frac{\partial^m \psi}{\partial x^m} \int dy e^{-i y p} y^{n+m}=\\
\\
=\left. \exp \left[ \frac{i \hbar}{2} \frac{\partial}{\partial p} \left( \frac{\partial}{\partial x} - \frac{\partial}{\partial x'} \right) \right] \delta (p)
\psi^*(x',t) \psi (x,t) \right|_{x'=x}.
\end{array}
\end{equation}
If we substitute $\psi = R e^{\frac{i}{\hbar} S}$ in the previous equation, we obtain to order $\hbar^3$:
\begin{equation}
F^W (x,p,t) = R^2 (x,t) \delta \left( p - \frac{\partial S}{\partial x} \right) + \frac{\hbar^2}{4} \left[ \left( \frac{\partial R}{\partial x} \right)^2 - R \frac{ \partial^2 R}{\partial x^2} \right] \delta'' \left( p - \frac{\partial S}{\partial x} \right) + \frac{\hbar^2}{24} R^2 \frac{\partial^3 S}{\partial x^3} \delta''' \left( p - \frac{\partial S}{\partial x} \right) + {\cal O} (\hbar^4).
\end{equation}
Notice that odd powers of $\hbar$ do not appear in the previous expansion, because the Wigner function is real. We conclude that the Wigner function can be regarded as a $\hbar$-deformation of the Bohmian distribution. If we want the two to match, we have to impose successive corrections to vanish, i.e. $R \frac{\partial^2 R}{ \partial x^2} - \left(\frac{\partial R}{\partial x} \right)^2 =0$ and $\frac{\partial^3 S}{ \partial x^3}=0$.
There are other conclusions that can be drawn from the expansion (21) of the Wigner function:

\noindent
(i) If the corrections to the de Broglie-Bohm distribution in eq.(21) and the quantum potential in eq.(15) are negligible (this corresponds to a formal $\hbar \to 0$ limit), we conclude that:
\begin{equation}
F^W_{cl} (x,p) = R^2_{cl} (x;t) \delta \left( p - \frac{\partial S_{cl}}{\partial x} (x;t) \right),
\end{equation}
where $R_{cl}$ and $S_{cl}$ obey the classical continuity and Hamilton-Jacobi equations, respectively. Consequently, this $F^W_{cl}$ is positive defined and obeys the classical Liouville equation. The expression on the right-hand side of eq.(22) is usually regarded as the classical limit of a quantum mechanical pure state \cite{Holland}.

\noindent
(ii) In most cases two distinct quasi-distributions can be expressed as $\hbar$-deformations of each other. Equation (21) could then be a hint that the de Broglie-Bohm distribution (16) is actually just another quasi-distribution. Indeed, we proved in a previous work that it can be implemented in quantum phase-space with a particular choice of Cohen's $f$-function \cite{Dias3}.

\noindent
(iii) Perhaps it is possible to re-express the $\hbar$-expansion (21) in terms of the star product $*_W$.

\vspace{0.2 cm}
\noindent
Indeed, the conjecture stated in (iii) is a particular case of the following, more general theorem:

\noindent
{\underline{\bf Theorem:}} Consider any of Cohen's functions $f ( \xi, \eta)$, such that:
\begin{equation}
f ( \xi, 0) = 1, \qquad \frac{\partial f}{\partial \eta} (0,0)=0.
\end{equation}
Then the following formula holds:
\begin{equation}
F^f (x,p,t) = R(x,t) *_f' \delta^f_{*'} \left( p - \frac{\partial S}{\partial x} (x,t) \right) *_f' R (x,t),
\end{equation}
where $R (x,t)$ and $S(x,t)$ are the solutions of the Bohm equations (14).$_{\Box}$.

\noindent
The proof of the theorem can be found in the appendix. The previous formula constitutes an alternative (and, hopefully, more intuitive) formulation of Cohen's classification of quasi-distributions, which can be stated as follows. Consider an arbitrary pair of real and continuous functions $R$ and $S$ satisfying the usual requirements of the causal interpretation \cite{Holland}. Given an arbitrary non-commutative associative $*'$-product, such that $A(x) *' B(x)= A(x) B(x)$, then the corresponding pure state quasi-distribution associated with $\psi = R e^{\frac{i}{\hbar}S}$ is given by equation (24).

Furthermore, this formula provides (i) a reformulation of the $f$-dual-map for density matrices (i.e. given $R$ and $S$, we automatically have the pure state $f$-quasi-distribution (24) which is the $f$-dual-map of the corresponding density matrix.), (ii) a previously missing quantization procedure for distributions. The statement that quantization is formally just a substitution of the standard product by a non-local star-product has so far only applied to the observable's sector. Now its validity has been extended to the dual sector as well. In the de Broglie-Bohm formulation quantum and classical pure states differ only in that the Hamilton-Jacobi functions $S$ and $S_{cl}$ appearing in equations (16,22) obey Hamilton-Jacobi equations with or without quantum potential, respectively. Likewise, any quasi-distribution is a $*$-deformed quantization of a classical pure state plus a  replacement of $S_{cl}$ by $S$. Notice also that the de Broglie-Bohm quantization prescription is perfectly compatible with the deformation procedure. In fact, we can recover (16) from (24) for a particular choice of $f$-function and associated $*$-product \cite{Dias3}. Finally, in this context we also see that the concept of classical pure state is a common feature of the semiclassical limit of all quasi-distributions.

Moreover, the "quasi-causal" form of the quasi-distributions (24) provides an alternative formulation of the dynamics in terms of Bohmian trajectories. To obtain the time evolution we just have to solve eqs.(14) and substitute the solutions $R (x,t)$ and $S(x,t)$ in eq.(24). In our opinion this is a much more elegant approach. In the usual quasi-distribution formalism, time evolution is obtained by solving the $*$-deformed Liouville equation: $\frac{\partial F^f}{\partial t} = \frac{1}{i \hbar} \left[\tilde H^f, F^f \right]_f'$, which is typically an infinite order partial differential equation and displays a set of solutions that do not allow for an interpretation in terms of causal trajectories. Equation (24), on the other hand, casts the time evolution of a quasi-distribution as the solution of a system of two second order partial differential equations and allows for a straightforward interpretation in terms of "interfering" Bohmian trajectories. Moreover, expansions like the one in eq.(21), seem to be related to the topic of Wigner trajectories, which correspond to order-by-order quantum corrections to the classical Hamilton trajectories.

Finally, let us make a comment on the non-commutability of star products. Let us start with the de Broglie-Bohm distribution (16). It corresponds {\it de facto} to an abelian star product in equation (24). Under a gauge transformation we obtain say the Wigner function with the non-commutative Groenewold star product (9). What have we benefited from this gauge transformation? We have removed the wave function from the observable's sector \cite{Holland}, and the price to pay was to introduce a non-commutative star-product in the dual sector. So, apparently, the role of non-commutativity is to replace the action of the pilot wave on the observable's sector, i.e. on quantities such as kinetic energy or angular momentum.

\section{Example}

To illustrate the previous result (24), we consider the simple example of a free gaussian wave packet. At $t=0$ the wave function reads \cite{Holland}:
\begin{equation}
\psi_0 (x) = \left( 2 \pi \sigma_0^2 \right)^{- \frac{1}{4}} \exp \left[ - \frac{x^2}{4 \sigma_0^2} + \frac{i}{\hbar} p_0 x \right].
\end{equation}
After a lapse of time $t$, we get from the free Schr\"odinger equation:
\begin{equation}
\psi(x,t) = \left( 2 \pi s_t^2 \right)^{- \frac{1}{4}} \exp \left[ - \frac{(x- ut)^2}{4 \sigma_0 s_t} + \frac{i}{\hbar} p_0 \left( x -\frac{1}{2} ut \right) \right],
\end{equation}
where $s_t = \sigma_0 \left( 1 + \frac{i \hbar t}{2m \sigma_0^2} \right)$ and $u=\frac{p_0}{m}$. The amplitude and the phase function read:
\begin{equation}
\left\{
\begin{array}{l}
R(x,t) = \left( 2 \pi \sigma^2 \right)^{- \frac{1}{4}} \exp \left[ - \frac{(x- ut)^2}{4 \sigma^2}  \right],\\
\\
S(x,t) = - \frac{\hbar}{2} \arctan \left( \frac{\hbar t}{2m \sigma_0^2} \right) + p_0 \left( x -\frac{1}{2} ut \right)+ \frac{\hbar^2 t}{8 m \sigma_0^2 \sigma^2} (x- ut)^2,
\end{array}
\right.
\end{equation}
where $\sigma \equiv | s_t|$. They solve the de Broglie-Bohm system (14). Substituting the previous expression for $S(x,t)$ in eq.(24) with $f= f_W =1$, we obtain:
\begin{equation}
F^W (x,p,t) = R(x,t) *_W \delta_*^W \left[ p -p _0 - \frac{\hbar^2t}{4m \sigma_0^2 \sigma^2} (x- ut) \right]*_W R(x , t).
\end{equation}
Given the fact that $\frac{\partial S}{\partial x}$ is (in this case) a linear function of $x$, we conclude from (11) that the $*$-delta function reduces to the ordinary delta function. Furthermore, if we execute the first $*$-product, we get:
$$
\begin{array}{c}
F^W (x,p,t)= \left\{ \sum_{n=0}^{\infty} \frac{1}{n!} \left( \frac{i \hbar}{2} \right)^n \frac{\partial^n R}{\partial x^n} \delta^{(n)} \left[ p -p _0 - \frac{\hbar^2t}{4m \sigma_0^2 \sigma^2} (x- ut) \right] \right\}*_W R(x , t)=\\
\\
= \left\{ \frac{1}{2 \pi} \sum_{n=0}^{\infty} \frac{1}{n!} \left( \frac{i \hbar}{2} \right)^n \frac{\partial^n R}{\partial x^n} \int dy (i y)^n \exp \left[iy \left( p -p _0 - \frac{\hbar^2t}{4m \sigma_0^2 \sigma^2} (x- ut)\right) \right] \right\}*_W R(x , t) = \\
\\
= \left\{ \frac{1}{2 \pi} \int dy  R \left( x -\frac{\hbar}{2} y , t \right) \exp \left[ iy \left(p -p _0 - \frac{\hbar^2t}{4m \sigma_0^2 \sigma^2} (x- ut)\right) \right] \right\}*_W R(x , t).
\end{array}
$$
A similar procedure for the second $*$-product yields:
\begin{equation}
\begin{array}{c}
F^W (x,p,t)= \frac{1}{2 \pi}  \sum_{n=0}^{\infty} \frac{1}{n!} \left(- \frac{i \hbar}{2} \right)^n \int dy  R \left( x -\frac{\hbar}{2} y , t \right) (iy)^n \frac{\partial^n R}{\partial x^n} \exp \left[ iy \left( p -p _0 - \frac{\hbar^2t}{4m \sigma_0^2 \sigma^2} (x- ut) \right) \right]=\\
\\
= \frac{1}{2 \pi}  \int dy  R \left( x -\frac{\hbar}{2} y , t \right) R \left( x + \frac{\hbar}{2} y , t \right) \exp \left[ iy \left( p -p _0 - \frac{\hbar^2t}{4m \sigma_0^2 \sigma^2} (x- ut) \right) \right]
\end{array}
\end{equation}
A brief calculation shows that (cf.(27)):$\frac{i}{\hbar} \left[ S \left( x  + \frac{\hbar}{2} y , t \right) - S \left( x -\frac{\hbar}{2} y , t \right) \right] =  iy \left(p _0 + \frac{\hbar^2t}{4m \sigma_0^2 \sigma^2} (x- ut)\right) $. Upon substitution of the previous expression in equation (29), we recover (17).

\section{Appendix}

To derive eq.(24), let us first prove the ensuing lemma. Let us start by considering the Mehta function \cite{Mehta},
\begin{equation}
F^S (x,p) = \frac{1}{\sqrt{2 \pi \hbar}} \psi^* (x) \phi (p) e^{\frac{i}{\hbar} xp},
\end{equation}
corresponding to $f_S (\xi , \eta) = e^{- \frac{i \hbar}{2} \xi \eta}$. This rule is associated with the {\it standard} ordering according to which all powers of the operator $\hat x$ precede all powers of the operator $\hat p$ \cite{Lee}. Here $\phi (p)$ stands for the Fourier transform of $\psi (x)$: $\phi (p) = \frac{1}{\sqrt{2 \pi \hbar}} \int dx \hspace{0.2 cm} \psi (x) e^{- \frac{i}{\hbar} xp}$. We now claim that the following lemma holds:
\\
{\underline{\bf Lemma:}} The quasi-distribution $F^S (x,p)$ can be written in the form:
\begin{equation}
F^S (x,p) = R (x) *_S' \delta^S_{*'} \left( p - \frac{\partial S}{\partial x} (x) \right) *_S' R (x),
\end{equation}
where\footnote{Notice that $*_S'$ is the dual star product of $*_{AS}$, since $f_{AS} (\xi, \eta)= e^{\frac{i \hbar}{2} \xi \eta} = f_S^{-1} (\xi, \eta)$. It corresponds to the anti-standard ordering according to which the $p$'s precede the $x$'s.}:
\begin{equation}
A(x,p) *_S' B(x,p) =  A(x,p) e^{i \hbar \frac{{\buildrel { \leftarrow}\over\partial}}{\partial p}  \frac{{\buildrel { \rightarrow}\over\partial}}{\partial x}} B(x,p)._{\Box}
\end{equation}
{\underline{\bf Proof of the lemma:}} Let us start by expressing $\psi (x)$ in the polar form (13):
\begin{equation}
F^S (x,p) = \frac{1}{2 \pi \hbar} R(x) e^{- \frac{i}{\hbar} S(x) } \int d x' \hspace{0.2 cm} R (x') e^{ \frac{i}{\hbar} S(x') - \frac{i}{\hbar} p (x'-x)}.
\end{equation}
On the other hand, if we expand $S(x')$ in powers of $(x'-x)$ we obtain:
\begin{equation}
\frac{i}{\hbar} \left\{ S(x') - p (x'-x) \right\} = \frac{i}{\hbar} \left\{ S(x) - \left( p - \frac{\partial S}{\partial x} (x) \right) (x'-x)  + \sum_{n=2}^{\infty} \frac{1}{n!} \frac{\partial^n S}{\partial x^n} (x) (x' -x)^n \right\} .
\end{equation}
Consequently:
\begin{equation}
\begin{array}{c}
F^S (x,p) = \frac{1}{2 \pi \hbar} R(x) \int dy \hspace{0.2 cm} R(x+y) e^{ - \frac{i}{\hbar}  y \left( p - \frac{\partial S}{\partial x} (x) \right)  + \frac{i}{\hbar} \sum_{n=2}^{\infty} \frac{1}{n!} \frac{\partial^n S}{\partial x^n} (x) y^n} =\\
\\
= R(x) *_S' \left\{  \frac{1}{2 \pi} \int dy \hspace{0.2 cm}  e^{ - i y  \left( p - \frac{\partial S}{\partial x} (x) \right)  + \frac{i}{\hbar} \sum_{n=2}^{\infty} \frac{1}{n!} \frac{\partial^n S}{\partial x^n} (x)(\hbar y)^n} \right\} *_S' R(x),
\end{array}
\end{equation}
where we expanded $R(x+y)$ in powers of $y$ and used the definition of $*_S'$ (eq.(32)). It remains to prove that the term in curly brackets is indeed equal to:
\begin{equation}
\delta^S_{*'} \left( p - \frac{\partial S}{\partial x} (x) \right) = \frac{1}{2 \pi} \int d y \hspace{0.2 cm} e_{*_S'}^{i y \left( p - \frac{\partial S}{\partial x} (x) \right)}.
\end{equation}
Let $A(x,p) = i y \left( p - \frac{\partial S}{\partial x} (x) \right)$. We then have: $ e_{*_S'}^{i y \left( p - \frac{\partial S}{\partial x} (x) \right)}= \sum_{n=0}^{\infty} \frac{1}{n!} \Omega_n$, where $\Omega_0 =1$ and $\Omega_{n+1} = A *_S' \Omega_n = \left( A - \hbar y \frac{\partial}{\partial x} \right) \Omega_n$. We conclude that: $e_{*_S'}^{i y \left( p - \frac{\partial S}{\partial x} (x) \right)}= e^{A - \hbar y \frac{\partial}{\partial x} } \Omega_0 = e^{iyp} \left( e^{\hat B + \hat C} \Omega_0 \right)$, where $\hat B = - i y \frac{\partial S}{\partial x} (x) $ and $\hat C = - \hbar y \frac{\partial}{\partial x}$. Now notice that: $\left[ \hat B , \hat C \right] = - i \hbar y^2
\frac{\partial^2 S}{\partial x^2}$. Moreover, any multiple commutator involving $\left[\hat B, \hat C \right]$ and any non zero number of $\hat B$'s vanishes, e.g. $\left[\left[ \hat B, \hat C \right], \hat B \right] = \left[\left[ \left[\hat B, \hat C \right], \hat C \right] , \hat B \right]= \cdots =0.$ The Baker-Campbel-Hausdorff formula \cite{Tannoudji} then reduces to:
\begin{equation}
e^{\hat B + \hat C} = \exp \left\{ \sum_{n=2}^{\infty} \frac{(-1)^n}{n!} \left[ \left[ \cdots \left[ \left[ \hat B , \hat C \right], \hat C \right] , \cdots , \hat C \right], \hat C \right] \right\} e^{\hat B} e^{\hat C},
\end{equation}
where for each $n$, the operator $\hat C$ appears $n-1$ times in the multiple commutator. A straightforward calculation yields: $\left[ \left[ \cdots \left[ \left[ \hat B , \hat C \right], \hat C \right] , \cdots , \hat C \right], \hat C \right] = - i \hbar^{n-1} y^n \frac{\partial^n S}{\partial x^n} (x)$. Consequently:
\begin{equation}
e_{*_S'}^{i y \left( p - \frac{\partial S}{\partial x} (x) \right)}= e^{iyp}  \exp \left\{ \frac{i}{\hbar} \sum_{n=2}^{\infty} \frac{(- \hbar y)^n}{n!} \frac{\partial^n S}{\partial x^n} (x) \right\} e^{-i y \frac{\partial S}{\partial x} (x)} e^{ - \hbar y \frac{\partial}{\partial x}} \Omega_0.
\end{equation}
Finally, taking into account that $e^{ - \hbar y \frac{\partial}{\partial x}} \Omega_0 =1$, we get:
\begin{equation}
\delta^S_{*'} \left( p - \frac{\partial S}{\partial x} (x) \right) = \frac{1}{2 \pi} \int dy \hspace{0.2 cm} e^{ iy  \left( p - \frac{\partial S}{\partial x} (x) \right) + \frac{i}{\hbar} \sum_{n=2}^{\infty} \frac{(-\hbar y)^n}{n!} \frac{\partial^n S}{\partial x^n} (x)},
\end{equation}
which is in perfect agreement with equation (35).$_{\Box}$

\noindent
As a consistency check, we can verify, using the previous equation, that: $p *_S' \delta^S_{*'} \left( p - \frac{\partial S}{\partial x} (x) \right) = \frac{\partial S}{\partial x} (x) *_S' \delta^S_{*'} \left( p - \frac{\partial S}{\partial x} (x) \right)$. We are now in a position to prove the theorem.
\\
{\underline{\bf Proof of the Theorem:}} Let us start by proving formula (24) for the time-independent case. First of all let us verify what are the implications of the constraints (23). For $f ( \xi, 0) =1$, it is easy to check that $F^f (x,p)$ admits the marginal distribution: $\int dp \hspace{0.2 cm} F^f (x,p) = | \psi (x)|^2 = R^2 (x)$. Moreover, this constraint entails that any $p$-independent function $g(x)$ remains unaltered under a change of function $f(\xi , \eta)$: $f \left( i \frac{\partial}{\partial x} , i \frac{\partial}{\partial p} \right) g(x) = f \left( i \frac{\partial}{\partial x},0 \right) g (x) = g(x)$. Finally, the second constraint (23) means that $p$ is also invariant under a change of $f(\xi , \eta)$: $
f \left( i \frac{\partial}{\partial x} , i \frac{\partial}{\partial p} \right) p = f \left( 0, i \frac{\partial}{\partial p} \right) p = f(0,0) p + i \frac{\partial f}{\partial \eta} (0,0) = p$. The sole purpose of these constraints is to leave $R(x)$, $\frac{\partial S}{\partial x} (x)$ and $p$ unchanged under a "gauge" transformation. The "gauge" transformation will thus only act on the star product. In particular, $f_W (\xi, \eta)$, $f_S (\xi, \eta)$ and  $f_B (\xi, \eta)$ \cite{Dias3} satisfy both constraints. From (8) we then have:
\begin{equation}
\begin{array}{c}
F^f (x,p) = f \left( i \frac{\partial}{\partial x} , i \frac{\partial}{\partial p} \right) f_S^{-1} \left( i \frac{\partial}{\partial x} , i \frac{\partial}{\partial p} \right)  F^S (x,p) = \\
\\
= f \left( i \frac{\partial}{\partial x} , i \frac{\partial}{\partial p} \right) f_S^{-1} \left( i \frac{\partial}{\partial x} , i \frac{\partial}{\partial p} \right) \left[ R (x) *_S' \delta^S_{*'} \left( p - \frac{\partial S}{\partial x} (x) \right) *_S' R (x) \right].
\end{array}
\end{equation}
But, from (8) and (23) we obtain:
\begin{equation}
\begin{array}{c}
F^f (x,p) = f \left( i \frac{\partial}{\partial x} , i \frac{\partial}{\partial p} \right) \left\{ f_S^{-1} \left( i \frac{\partial}{\partial x} , i \frac{\partial}{\partial p} \right)
\left[ R (x) *_S' \delta^S_{*'} \left( p - \frac{\partial S}{\partial x} (x) \right) *_S' R (x) \right] \right\}=\\
\\
= f \left( i \frac{\partial}{\partial x} , i \frac{\partial}{\partial p} \right) \left\{ R (x) *_W \delta^W_* \left( p - \frac{\partial S}{\partial x} (x) \right) *_W R (x) \right\} =  R (x) *_f' \delta^f_{*'} \left( p - \frac{\partial S}{\partial x} (x) \right) *_f' R (x).
\end{array}
\end{equation}
As a consistency check, let us verify that for $f= f_W$, the previous expression coincides up to order $\hbar^3$ with eq.(21). From equation (11), we get for $A(x,p) = p -\frac{\partial S}{\partial x} (x)$: $\Theta_1 (x,p) =0$, $\Theta_2 (x,p) = - \frac{\partial^3 S}{\partial x^3} (x)$. Consequently: $\delta^W_* \left( p - \frac{\partial S}{\partial x} \right) = \delta \left( p - \frac{\partial S}{\partial x} \right) + \frac{\hbar^2}{24} \frac{\partial^3 S}{\partial x^3} \delta''' \left( p - \frac{\partial S}{\partial x} \right) + {\cal O} (\hbar^4)$. Substituting in eq.(41), we do indeed recover (21). Moreover, notice that for $S(x)=a +b x +cx^2$, the $*$-delta function reduces to the ordinary delta function.

\noindent
To obtain the time-evolution of $F^f (x,p)$ (eq.(24)) we just have to follow all the previous steps, except that we now use the explicit time dependent polar form of the wavefunction $\psi (x,t) = R(x,t) \exp \left[\frac{i}{\hbar} S(x,t) \right]._{\Box}$

\vspace{1 cm}

\begin{center}

{\large{{\bf Acknowledgments}}}

\end{center}

\vspace{0.3 cm}
\noindent
We would like to thank Jo\~ao Marto for useful suggestions and for reading the manuscript. This work was partially supported by the grants ESO/PRO/1258/98 and CERN/P/Fis/15190/1999.

\end{document}